\documentclass[a4paper,11pt]{article}
\usepackage{pos}

\usepackage[separate-uncertainty=true]{siunitx}
\usepackage{enumitem}
\usepackage{lineno}

\usepackage{xspace}
\newcommand{\Xmax}{\ensuremath{X_\text{max}}\xspace}
\newcommand{\cor}[1]{\mbox{CORSIK\kern -0.05em A~#1}\xspace}

\title{Simulating radio emission from extensive air showers with \cor{8}}

\author*[a]{Marvin Gottowik}
\affiliation[a]{Karlsruhe Institute of Technology, Institute for Astroparticle Physics, Karlsruhe, Germany}

\onbehalf{for the \cor{8} collaboration\footnote[2]{email: \href{mailto:corsika8@kit.edu}{\texttt{corsika8@kit.edu}}}\newline\textnormal{(a complete list of authors can be found at the end of the proceedings)}}

\abstract{
\cor{8} is a modern, flexible framework for simulating particle cascades in air and dense media, allowing for fully customizable shower simulations. The radio module autonomously handles electric field calculations and propagation to observer locations. It supports simultaneous simulations with both the ``Endpoint formalism'' as implemented in CoREAS and the ``ZHS'' algorithm from ZHAireS. In this contribution, we validate the radio module by comparing air-shower simulations in \cor{8}, \cor{7}, and ZHAireS. We investigate the impact of simulation parameters, such as the step size of particle tracks, on the resulting radio signals and perform a detailed comparison of the ``Endpoints'' and ``ZHS'' formalisms. For the same underlying showers simulated with \cor{8} with optimized step sizes, both formalisms converge to the same radiation energy within \SI{2}{\percent}. For \cor{8} and \cor{7}, agreement on the radiation energy is better than \SI{10}{\percent} in the 30–80 MHz band and improves to better than \SI{2}{\percent} for 50–350 MHz. This consistency provides further confirmation of the accuracy of microscopic air-shower radio emission simulations which is crucial for precise energy scale estimations using radio detection.
}

\ConferenceLogo{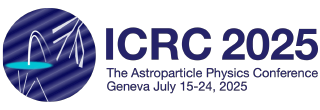}

\FullConference{39th International Cosmic Ray Conference (ICRC2025)\\
 15–24 July 2025\\
Geneva, Switzerland\\}

\begin{document}
\maketitle

\section{Introduction}
Radio detection of extensive air showers has proven to be a valuable technique in astroparticle physics, offering high-precision measurements of cosmic-ray energy and shower development~\cite{Huege:2016veh}. As the field moves toward increasingly accurate and large-scale radio observatories, the need for reliable and flexible simulation tools becomes more critical. Simulating the radio emission from air showers and particle cascades in dense media requires a detailed treatment of particle interactions and electromagnetic radiation, along with the ability to model diverse detector configurations and environmental conditions. Modern simulation frameworks must therefore combine physical accuracy with modularity and extensibility to support a wide range of scientific goals.

\cor{8} is a modern, modular framework for the simulation of particle cascades in air and dense media~\cite{corsikaNG, corsika8_radio}. Its design allows for highly customizable simulations, integrating various physical models within a unified and extensible structure. The \cor{8} code is considered ``physics complete'' and a first expert version of it was released at the end of 2024~\cite{corsika8_beta}. An overview of the full capabilities of \cor{8} is presented in a separate contribution to this conference~\cite{corsika8_overview}. The radio module in \cor{8} autonomously handles the calculation and propagation of the electric field to user-defined observer locations. It supports both the ``Endpoint formalism'', as implemented in CoREAS~\cite{Huege:2013vt} (as part of \cor{7}~\cite{corsika}), and the ``ZHS algorithm'' known from ZHAireS~\cite{ALVAREZMUNIZ2012325}, allowing for direct comparisons within a consistent framework.

In this work, we present a validation of the \cor{8} radio module through detailed comparisons with established simulation tools, namely \cor{7} and ZHAireS. We study the influence of technical steering parameters, such as step sizes used in the particle tracking, quantified in terms of the maximum angular deflection a particle is allowed to undergo between successive tracking points, on the resulting radio signal and investigate the consistency between the CoREAS-like endpoint algorithm and ZHS formalisms when applied to the same underlying shower. A key observable in this context is the radiation energy, defined as the total energy deposited on the ground by radio waves in a given frequency band. It plays a crucial role in energy estimation for cosmic-ray air showers~\cite{Huege:2016veh} and is therefore an important benchmark for validating the physical accuracy of radio emission simulations. The results discussed here summarize the key findings on radio emission simulations with \cor{8} published in Ref.~\cite{corsika8_radio}, providing an overview of its capabilities and validation status. In addition to the validation presented here, a separate contribution to this conference highlights the unique capabilities of \cor{8} to simulate radio emission and signal propagation in complex and inhomogeneous media, such as realistic glacial environments relevant for neutrino observatories~\cite{corsika8_eisvogel}.

\section{Design of the radio module}

\begin{figure}
    \centering
    \includegraphics[width=\linewidth]{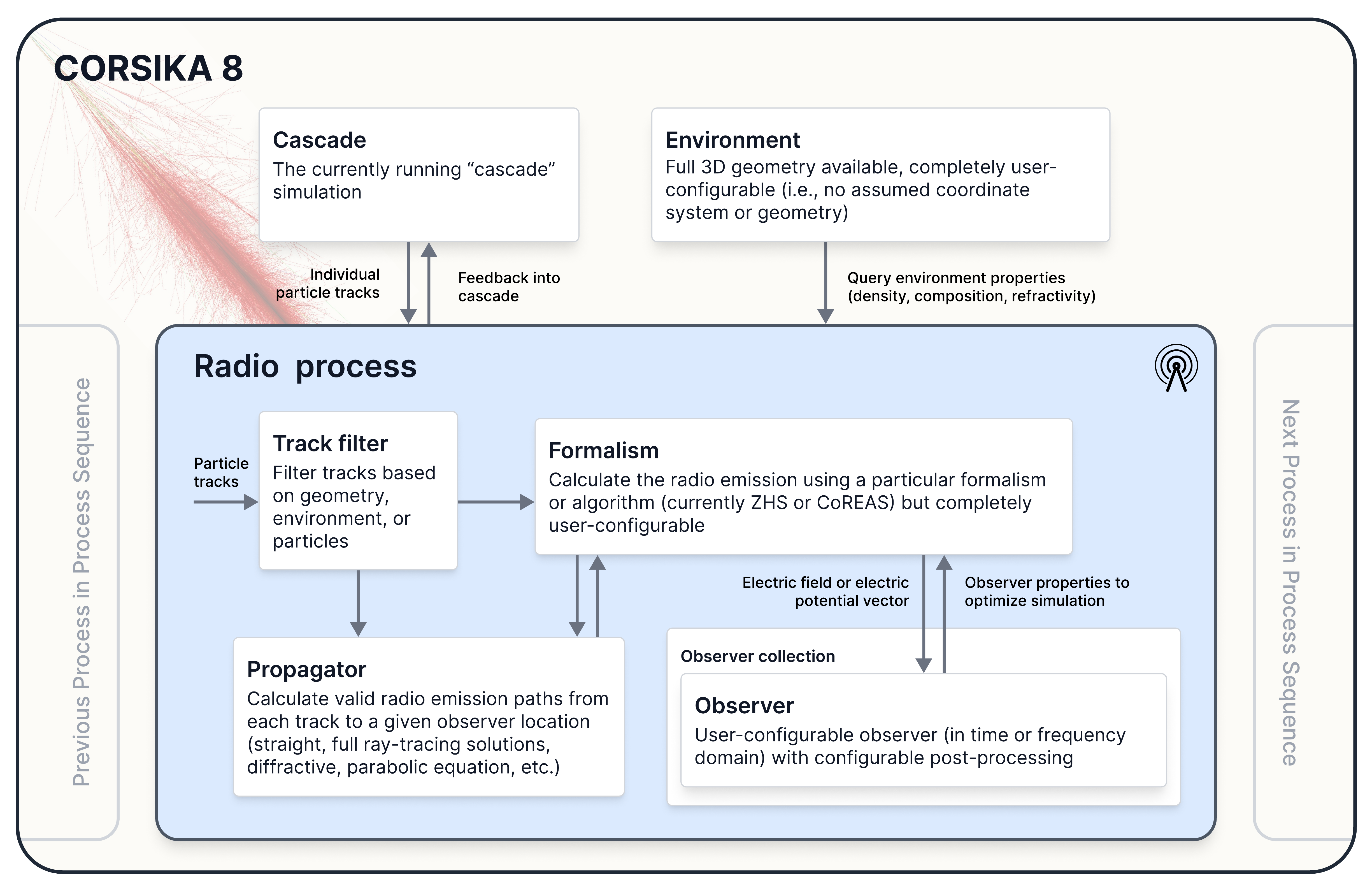}
    \caption{Schematic overview of the radio process currently implemented in \cor{8}, and showing how it integrates with the framework.}
    \label{fig:radio_arch}
\end{figure}

The \cor{8} radio module is designed with the core principles of the framework in mind: modularity, extensibility, and physical accuracy. As illustrated in Figure~\ref{fig:radio_arch}, it integrates seamlessly with the simulation’s main loop (\textit{Cascade}) and the user-defined \textit{Environment}, which provides spatially dependent physical properties such as refractive index, density, and magnetic field. This enables fully self-consistent simulations of radio emission in realistic atmospheric and detector conditions.

Each \textit{Radio Process} consists of interchangeable components, designed to be configurable and replaceable with user-defined C++ implementations. This makes the module highly adaptable to various use cases. The main components are:
\begin{itemize}
\item \textbf{Track Filter}: Determines which particle tracks contribute to the radio signal. Typically, only electrons and positrons are considered, but filtering can also be applied by energy, depth, or other user-defined criteria.
\item \textbf{Formalism}: Computes the electric field based on the selected particle tracks. Currently, two independent time-domain formalisms are supported: the ``CoREAS'' endpoint formalism and the ``ZHS'' algorithm. Both are implemented in a way that closely mirrors their original versions, allowing direct comparison with \cor{7} and ZHAireS.
\item \textbf{Propagator}: Calculates the propagation of the radio signal from its emission point to each observer, including travel time and path. Three propagators are available: a general-purpose ray-integrator for complex media, and two faster approximations for stratified atmospheric models (uniform and exponential). All use straight-line paths for now, but the design allows for future integration of ray-traced or curved-path models.
\item \textbf{Observer}: Handles the reception and storage of the electric field at each defined observer location. Observers are organized into collections and support configurable parameters such as sampling rate and time window.
\end{itemize}

This modular design allows users to tailor the radio simulation pipeline to a wide range of scientific goals. The clear separation between the responsibilities of different components also facilitates the development and integration of new features. A distinctive feature of \cor{8} is its ability to run multiple \textit{Radio Processes} simultaneously on the exact same air shower. Thus, different formalisms can be applied in parallel using the same underlying shower development, including identical particle tracks, interaction history, and random numbers. As a result, differences in the predicted radio signals can be attributed solely to the emission model itself, rather than to shower-to-shower fluctuations or implementation details in other parts of the simulation. This capability enables true, unbiased comparisons of radio emission formalisms within a unified, consistent framework.

\section{Validation of the radio module}

\cor{8} follows strict development guidelines across all modules, emphasizing physical correctness, code reliability, and extensive test coverage. The radio module is no exception: in addition to unit tests with high code coverage, it underwent validation using a well-understood physical scenario, a single relativistic electron in a uniform magnetic field. We consider an electron with energy $\sim$\SI{11.4}{MeV} ($\beta = 0.999$) performing circular motion in vacuum ($n = 1$) under the influence of a magnetic field aligned along the $z$-axis. The track radius is $L = \SI{100}{m}$, and the motion is confined to the $x$–$y$ plane. An observer is placed in the same plane at a distance of \SI{30}{km} from the center of the circle to ensure far-field conditions. 

The simulation uses \cor{8}’s native ``leapfrog'' tracking algorithm to determine the particle trajectory in the magnetic field. To ensure sufficient resolution for calculating the broadband synchrotron pulse, the tracking is configured to introduce a new step after each geomagnetic deflection of $10^{-4}$\,rad. The electron is followed for exactly one full revolution. The associated synchrotron radiation is calculated using both the ``CoREAS'' (endpoint with ZHS fallback close to the Cherenkov ring) and ``ZHS'' formalisms implemented in the radio module. To validate the results, we compare them to an analytical solution as described in Ref.~\cite{James:2011}, also used in previous validation efforts.

\begin{figure}
    \centering
    \includegraphics[height=5.2cm]{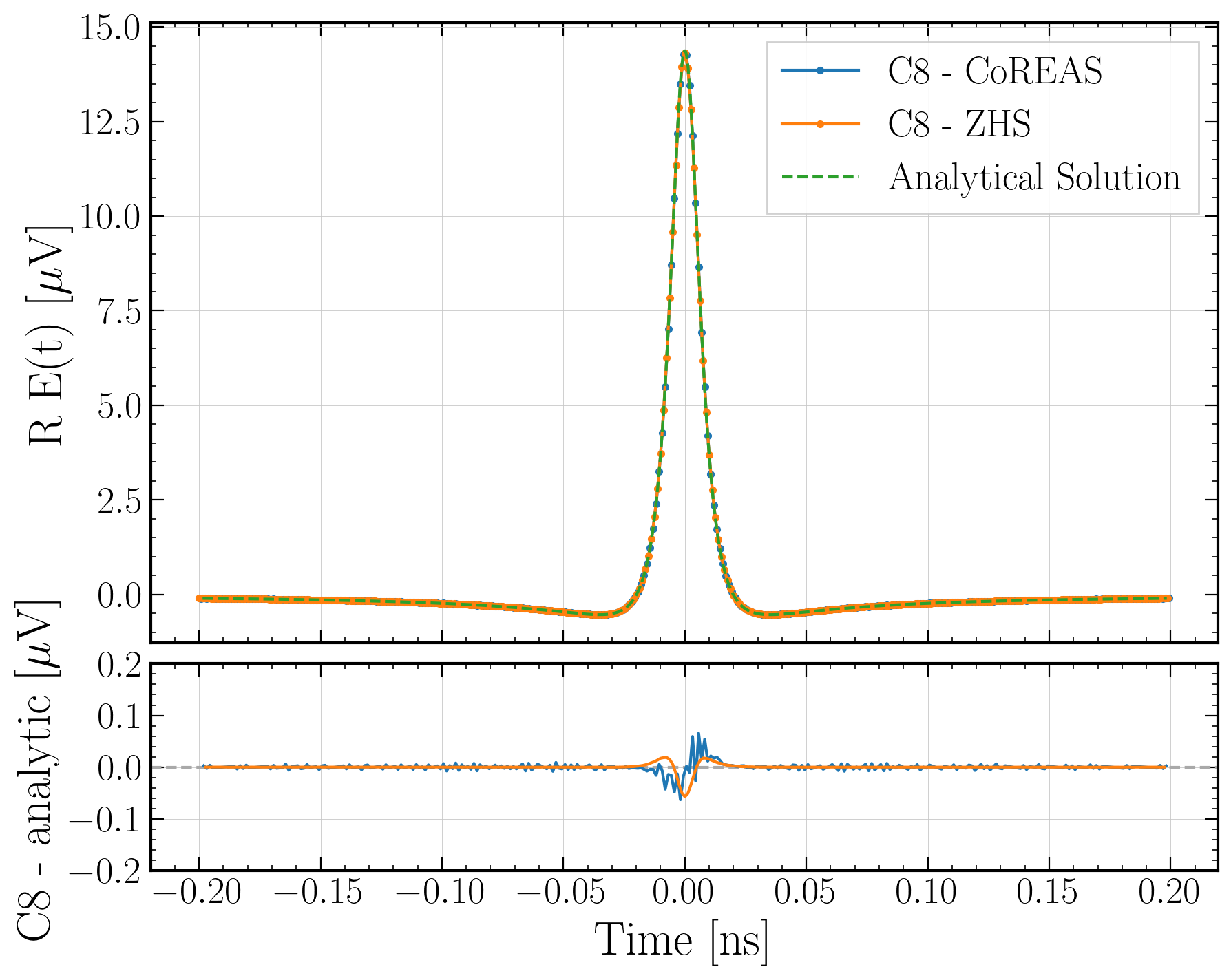}%
    \hspace{0.5cm}%
    \includegraphics[height=5.2cm]{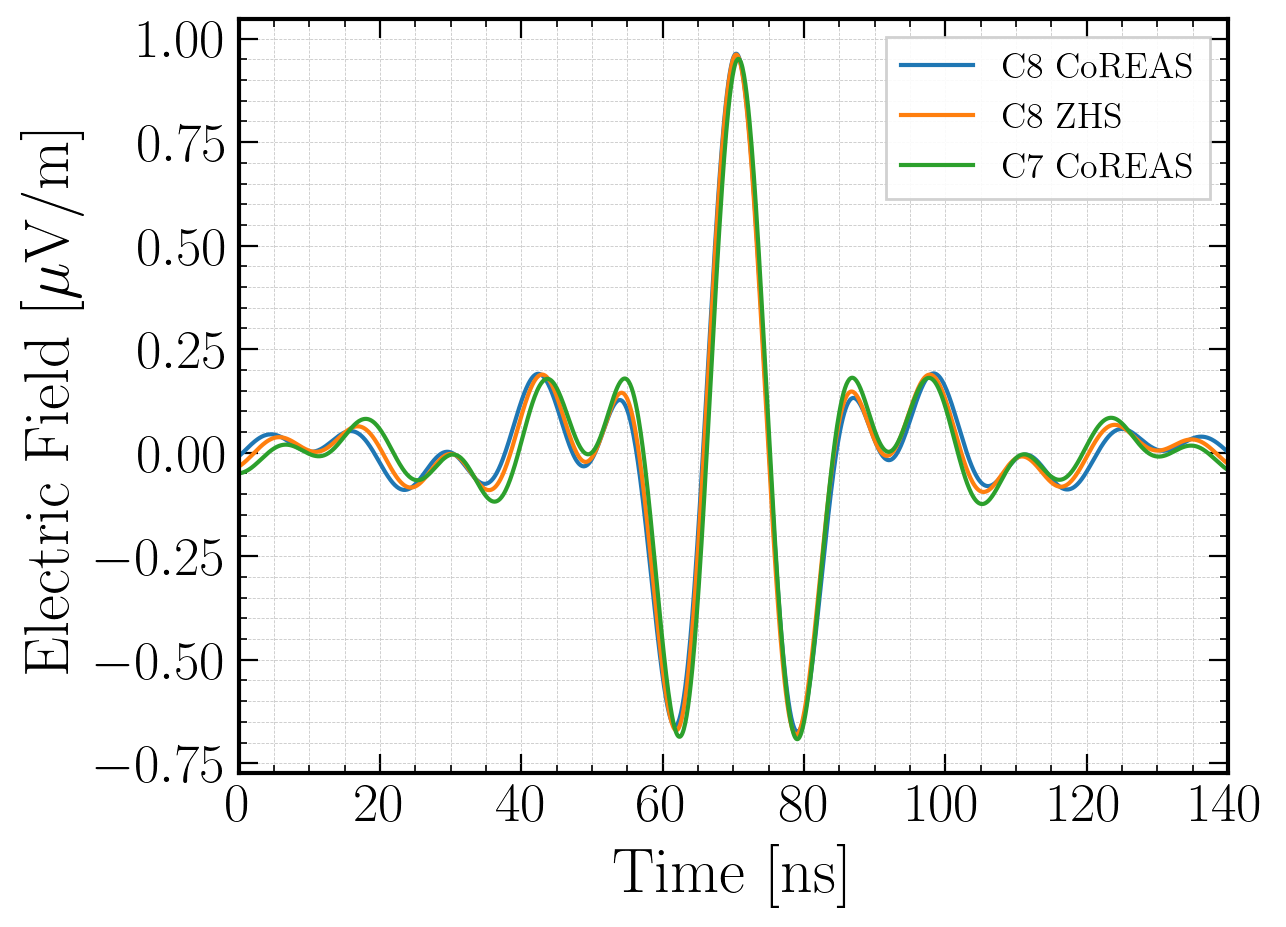}
    \caption{Left: Radio pulse from a single electron moving in a circular orbit in a uniform magnetic field. Shown are results from the native particle tracking in \cor{8} (C8) using a very fine step size and an analytical solution. Right: Comparison of radio pulses predicted by \cor{8} using the ``CoREAS'' and ``ZHS'' formalisms, and by \cor{7} (C7), for comparable showers with similar longitudinal profiles observed at \SI{100}{m} from the shower axis near the Cherenkov ring. The traces represent the geomagnetic component, filtered to the \SIrange{30}{80}{MHz} band.}
    \label{fig:valid_pulses}
\end{figure}

Figure~\ref{fig:valid_pulses} (left) shows the resulting electric field pulse in the $y$-polarization. Both formalisms produce results that agree with the analytical solution to within approximately \SI{1}{\percent}. Minor deviations near the peak are attributed to finite time binning in the simulation. Small oscillations observed in the CoREAS-style result stem from interference between the internal step timing and the output sampling interval and are expected to average out in realistic air-shower scenarios.

This test confirms that both radio-emission formalisms in \cor{8} yield physically accurate results when used with the native tracking algorithm, and that the observer handling and electric field propagation are correctly implemented.

\section{Air Shower comparison}

To evaluate the performance of the \cor{8} radio module under realistic conditions, we simulated a vertical \SI{100}{PeV} iron-induced air shower and compared the resulting radio emission to predictions from \cor{7} and ZHAireS. All simulations were configured with consistent technical parameters, including identical hadronic interaction models, thinning levels, cut energies for different particle types, and a common atmospheric and geomagnetic field setup. Particle tracking was left at standard values in \cor{7} (\texttt{STEPFC} = 1, maximum geomagnetic deflection per step = 0.2\,rad) and ZHAireS. For \cor{8}, the maximum allowed geomagnetic deflection before adding another tracking step was also set to 0.2 rad.

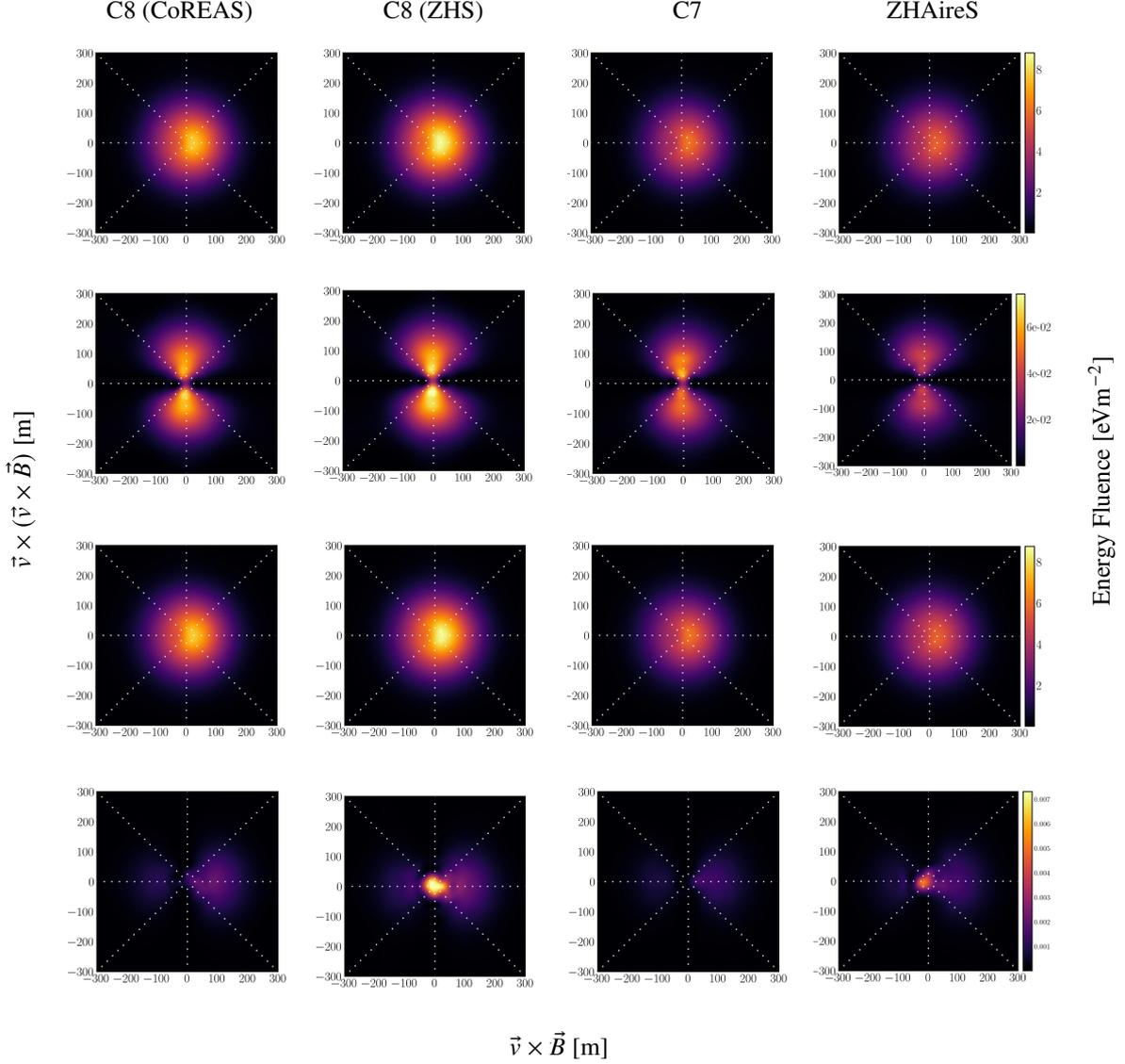
\begin{figure*}
    \centering
    \includegraphics[width=0.99\linewidth, trim={3.7cm 4.45cm 3.2cm 4cm}, clip]{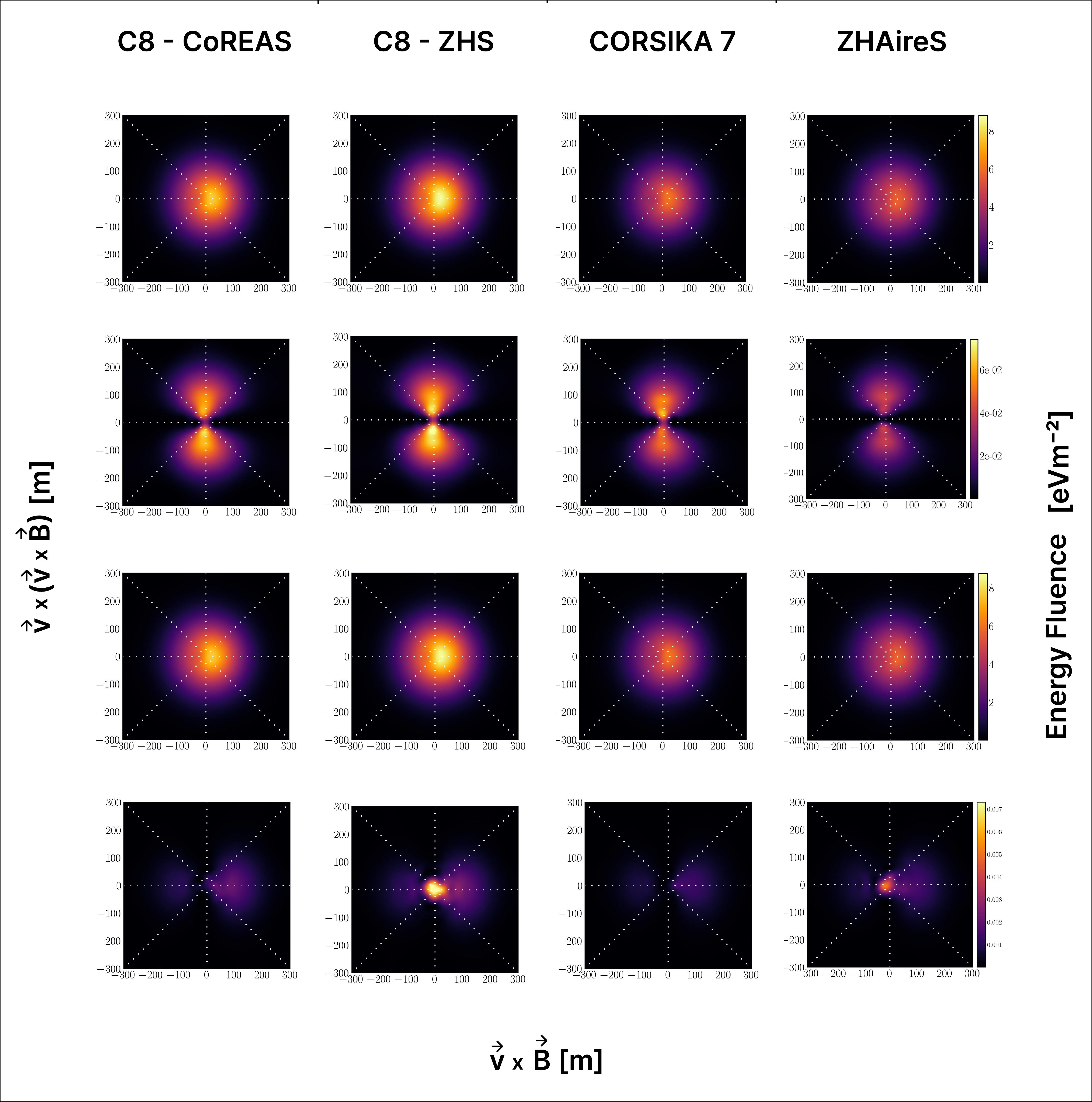}

    \begin{picture}(0,0)
      \put(-180, 425){\small C8 (CoREAS)}
      \put(-70, 425){\small C8 (ZHS)}
      \put(45, 425){\small C7}
      \put(130, 425){\small ZHAireS}
        
      \put(-20, 10){\small $\vec{v} \times \vec{B}$ [m]}
        
      \put(-220, 205){\rotatebox{90}{\small $\vec{v} \times (\vec{v} \times \vec{B})$ [m]}}
        
      \put(210, 190){\rotatebox{90}{\small Energy Fluence [eVm$^{-2}$]}}
    \end{picture}
    
    \caption{Energy fluence maps for different signal polarizations of the electric field in the \SIrange{30}{80}{\mega\hertz} frequency band for \cor{8} (C8) with ``CoREAS'' and ``ZHS'' formalism, \cor{7} (C7) with CoREAS extension, and ZHAireS. The order of the polarizations we see starting from top to bottom is: absolute value, $\vec{v} \times (\vec{v} \times \vec{B})$, $\vec{v} \times \vec{B}$, and $\vec{v}$, where $\vec{v}$ corresponds to the shower axis and $\vec{B}$ corresponds to the magnetic field axis. Please note the strongly different z-axis scales.}
    \label{fig:maps}
\end{figure*}

For\hspace{0.3em}\cor{8}, both emission formalisms were applied simultaneously to the same underlying shower. All three codes produced similar longitudinal shower profiles, with differences in \Xmax limited to about \SI{10}{g/cm^2}. Two-dimensional maps of the radio energy fluence in the \SIrange{30}{80}{MHz} band are shown in Fig. \ref{fig:maps}. We obtain good agreement across all codes in terms of shape, symmetry, and overall emission footprint. However, \cor{8} systematically predicts a $\sim\SI{25}{\percent}$ higher fluence than \cor{7} and ZHAireS. This initial difference arises from coarse tracking settings and should not be interpreted as a systematic discrepancy. We find that this difference decreases significantly when finer step sizes are used in particle tracking. Such a dependence was already found in an earlier study for \cor{7}~\cite{Gottowik:2018unc}. A similar plot with optimized step sizes for \cor{7} and \cor{8} can be found in Ref.~\cite{Gaudu:2024status}.

\cor{8} allows for direct comparison of the radio emission of ``CoREAS'' and ``ZHS'' formalism for identical showers. This approach removes variations in shower development that would otherwise arise when comparing across different codes. Using 100 vertical air showers simulated with optimized step sizes of 0.001\,rad in \cor{8} and \cor{7}, we compared the resulting radio pulses and average radiation energies in a consistent setting. The point of first interaction was fixed in all simulations to minimize shower-to-shower fluctuations. Radio pulses simulated at an observer distance of \SI{100}{m} from the shower axis, near the Cherenkov angle, show excellent agreement between \cor{8} and \cor{7}, as well as between the two formalisms within \cor{8}, as illustrated in Figure~\ref{fig:valid_pulses} (right). These results are based on independent but comparable showers simulated with both codes that yield similar longitudinal profiles. The average radiation energy predicted by both formalism in \cor{8} agree within \SI{2}{\percent} in the \SIrange{30}{80}{MHz} band and within \SI{1}{\percent} in the \SIrange{50}{350}{MHz} band. A comparison of the average radiation energy from the ``CoREAS'' implementations in \cor{8} and \cor{7} yields agreement within \SI{10}{\percent} in the \SIrange{30}{80}{MHz} band and \SI{2}{\percent} in the \SIrange{50}{350}{MHz} band.

\section{Summary}
\cor{8} can simulate radio emission from both extensive air showers and particle cascades in dense media. We have successfully validated its radio module through detailed comparisons with established simulation tools, \cor{7} and ZHAireS, demonstrating consistency in key observables such as radiation energy across a range of frequencies. The ability to simulate the radio emission both with the Endpoint and ZHS formalisms within a unified framework, and the close agreement for simulations with a small step size, further reinforces the robustness of the microscopic approach to radio-emission modeling. This proceeding highlights the key validation results of the radio module in \cor{8}. A more detailed discussion, including a systematic parameter scan demonstrating how the step-size setting affects the predicted radiation energy, as well as quantitative comparisons of the spatial distribution of the radio emission in the shower plane, can be found in the corresponding journal publication~\cite{corsika8_radio}. These results confirm that \cor{8} provides a reliable and flexible platform for high-precision radio simulations, supporting current and future efforts in radio-based astroparticle physics.

\clearpage

\section*{The \cor{8} Collaboration}

\begin{sloppypar}\noindent
J.M.~Alameddine$^{1,2}$,
J.~Albrecht$^{1,2}$,
A.A.~Alves Jr.$^{3,4}$,
J.~Ammerman-Yebra$^{5}$,
L.~Arrabito$^{6}$,
D.~Baack$^{1,2}$,
R.~Cesista$^{6}$,
A.~Coleman$^{7}$,
C.~Deaconu$^{8}$,
H.~Dembinski$^{1,2}$,
D.~Elsässer$^{1,2}$,
R.~Engel$^{3}$,
A.~Faure$^{6}$,
A.~Ferrari$^{3}$,
C.~Gaudu$^{9}$,
C.~Glaser$^{7,1}$,
M.~Gottowik$^{3}$,
D.~Heck$^{3}$,
T.~Huege$^{3,10}$,
K.H.~Kampert$^{9}$,
N.~Karastathis$^{3}$,
L.~Nellen$^{11}$,
D.~Parello$^{12,13}$,
T.~Pierog$^{3}$,
R.~Prechelt$^{14}$,
M.~Reininghaus$^{15}$,
W.~Rhode$^{1,2}$,
F.~Riehn$^{1}$,
M.~Sackel$^{1,2}$,
P.~Sampathkumar$^{3}$,
A.~Sandrock$^{9}$,
J.~Soedingrekso$^{1,2}$,
R.~Ulrich$^{3}$,
P.~Windischhofer$^{8}$,
B.~Yue$^{9}$

\vspace{1ex}
\begin{center}
\rule{0.1\columnwidth}{0.5pt}
\raisebox{-0.4ex}{\scriptsize$\bullet$}
\rule{0.1\columnwidth}{0.5pt}
\end{center}
\vspace{1ex}

\begin{itemize}[labelsep=0.2em,align=right,labelwidth=0.7em,labelindent=0em,leftmargin=2em,noitemsep,before={\renewcommand\makelabel[1]{##1 }}]
\item[$^{1}$] Technische Universität Dortmund (TU), Department of Physics, Dortmund, Germany
\item[$^{2}$] Lamarr Institute for Machine Learning and Artificial Intelligence, Dortmund, Germany
\item[$^{3}$] Karlsruhe Institute of Technology (KIT), Institute for Astroparticle Physics (IAP), Karlsruhe, Germany
\item[$^{4}$] University of Cincinnati, Cincinnati, OH, United States
\item[$^{5}$] IMAPP, Radboud University Nijmegen, Nijmegen, The Netherlands
\item[$^{6}$] Laboratoire Univers \& Particules de Montpellier, CNRS \& Université de Montpellier (UMR-5299), 34095 Montpellier, France
\item[$^{7}$] Uppsala University, Department of Physics and Astronomy, Uppsala, Sweden
\item[$^{8}$] Department of Physics, Enrico Fermi Institute, Kavli Institute for Cosmological Physics, University of Chicago, Chicago, IL 60637, USA
\item[$^{9}$] Bergische Universität Wuppertal, Department of Physics, Wuppertal, Germany
\item[$^{10}$] Vrije Universiteit Brussel, Astrophysical Institute, Brussels, Belgium
\item[$^{11}$] Universidad Nacional Autónoma de México (UNAM), Instituto de Ciencias Nucleares, México, México
\item[$^{12}$] DALI, Univ Perpignan, Perpignan, France
\item[$^{13}$] LIRMM Univ Montpellier, CNRS, Montpellier, France
\item[$^{14}$] University of Hawai'i at Manoa, Department of Physics and Astronomy, Honolulu, USA
\item[$^{15}$] Independent researcher
\end{itemize}

\end{sloppypar}

\section*{Acknowledgements}
\noindent
We thank T.~Sjöstrand, L.~Lönnblad, and the Pythia 8 collaborators for their support in the
implementation of Pythia 8/Angantyr in CORSIKA 8.

This research was funded by the Deutsche Forschungsgemeinschaft (DFG, German Research Foundation) – Projektnummer 445154105 and Collaborative Research Center SFB1491 "Cosmic Interacting Matters - From Source to Signal". This research has been partially funded by the Federal Ministry of Education and research of Germany and the state of North Rhine-Westphalia as part of the Lamarr Institute for Machine Learning and Artificial Intelligence. We acknowledge support through project UNAM-PAPIIT IN114924.

This work has also received financial support from Ministerio de Ciencia e Innovación/Agencia Estatal de Investigación (PRE2020-092276). A.~Coleman is supported by the Swedish Research Council (Vetenskapsrådet) under project no. 2021-05449. C.~Glaser is supported by the Swedish Research Council (Vetenskapsrådet) under project no. 2021-05449 and the European Union. F.~Riehn received funding from the European Union’s Horizon 2020 research and innovation programme under the Marie Skłodowska-Curie grant agreement No.~101065027. P.~Windischhofer and C.~Deaconu thank the NSF for Award 2411662. The authors acknowledge support by the High Performance and Cloud Computing Group at the Zentrum für Datenverarbeitung of the University of Tübingen, the state of Baden-Württemberg through bwHPC and the German Research Foundation (DFG) through grant no INST 37/935-1 FUGG. The computations were partially carried out on the PLEIADES cluster at the University of Wuppertal, which was supported by the Deutsche Forschungsgemeinschaft (DFG, grant No. INST 218/78-1 FUGG) and the Bundesministerium für Bildung und Forschung (BMBF).


\begin{thebibliography}{99}
\def\vyp#1#2#3{\textbf{#1} (#2) #3} 

\bibitem{Huege:2016veh}
T. Huege, Phys. Rept. \vyp{1}{2016}{620}.

\bibitem{corsikaNG}
R. Engel \emph{et al.} (\cor{8} Collaboration), Comput. Softw. Big Sci. \vyp{3}{2019}{2}.

\bibitem{corsika8_radio}
J.M. Alameddine \emph{et al.} (\cor{8} Collaboration), Astropart. Phys. \vyp{166}{2025}{103072}.

\bibitem{corsika8_beta}
\cor{8} beta release (v1.0-beta1), Zenodo, \url{https://doi.org/10.5281/zenodo.15525792}.

\bibitem{corsika8_overview}
F. Riehn \emph{et al.} (\cor{8} Collaboration), PoS(ICRC2025)371.

\bibitem{Huege:2013vt}
T. Huege, M. Ludwig, and C. W. James, AIP Conf. Proc. 1535, 128 (2013).

\bibitem{corsika}
D. Heck \emph{et al.}, FZKA Tech. Umw. Wis. B 6019, (1998).

\bibitem{ALVAREZMUNIZ2012325}
J. Alvarez-Muñiz, W. R. Carvalho, and E. Zas, Astropart. Phys. \vyp{35}{2012}{325--341}.

\bibitem{corsika8_eisvogel}
P. Windischhofer \emph{et al.} (\cor{8} Collaboration), PoS(ICRC2025)983.

\bibitem{James:2011}
C. W. James \emph{et al.}, Phys. Rev. E \vyp{84}{2011}{056602}.

\bibitem{Gottowik:2018unc}
M. Gottowik \emph{et al.}, Astropart. Phys. \vyp{103}{2018}{87}.

\bibitem{Gaudu:2024status}
C. Gaudu (\cor{8} Collaboration), arXiv:2412.15097 [astro-ph.HE].

\end{thebibliography}
\end{document}